%% file: main.tex
\documentclass[sigconf]{acmart}

\AtBeginDocument{%
  \providecommand\BibTeX{{%
    \normalfont B\kern-0.5em{\scshape i\kern-0.25em b}\kern-0.8em\TeX}}}
\usepackage{balance}
\usepackage[ruled]{algorithm2e}
\usepackage{multirow}
\usepackage{subfigure}
\usepackage{makecell}
\usepackage{bbm}
\usepackage{pgfplots}
\usetikzlibrary{patterns}
\usepackage{filecontents}
\usepackage{sistyle}
\SIthousandsep{,}
\usepackage{enumitem}
\usepackage{xcolor}

\definecolor{c1}{RGB}{255,255,157}
\definecolor{c2}{RGB}{190, 235, 159}
\definecolor{c3}{RGB}{0,163,136}

\definecolor{cgreen}{rgb}{0.0, 0.62, 0.42}

\copyrightyear{2021}
\acmYear{2021}
\setcopyright{acmcopyright}\acmConference[SIGIR '21]{Proceedings of the 44th
International ACM SIGIR Conference on Research and Development in Information
Retrieval}{July 11--15, 2021}{Virtual Event, Canada}
\acmBooktitle{Proceedings of the 44th International ACM SIGIR Conference on
Research and Development in Information Retrieval (SIGIR '21), July 11--15, 2021,
Virtual Event, Canada}
\acmPrice{15.00}
\acmDOI{10.1145/3404835.3462869}
\acmISBN{978-1-4503-8037-9/21/07}

\begin{document}
\fancyhead{}
\title{B-PROP: Bootstrapped Pre-training with Representative Words Prediction for Ad-hoc Retrieval}
\author{Xinyu Ma, Jiafeng Guo, Ruqing Zhang, Yixing Fan, Yingyan Li and Xueqi Cheng}
\affiliation{
  \institution{
    CAS Key Lab of Network Data Science and Technology, Institute of Computing Technology, \\ Chinese Academy of Sciences, Beijing, China\\
    University of Chinese Academy of Sciences, Beijing, China\\}
    \country{}
}
\email{{maxinyu17g,guojiafeng,zhangruqing,fanyixing,liyingyan19s,cxq}@ict.ac.cn}


\begin{abstract}
Pre-training and fine-tuning have achieved remarkable success in many downstream natural language processing (NLP) tasks.
Recently, pre-training methods tailored for information retrieval (IR) have also been explored, and the latest success is the PROP method which has reached new SOTA on a variety of ad-hoc retrieval benchmarks. The basic idea of PROP is to construct the \textit{representative words prediction} (ROP) task for pre-training inspired by the query likelihood model. Despite its exciting performance, the effectiveness of PROP might be bounded by the classical unigram language model adopted in the ROP task construction process. To tackle this problem, we propose a bootstrapped pre-training method (namely B-PROP) based on BERT for ad-hoc retrieval. The key idea is to use the powerful contextual language model BERT to replace the classical unigram language model for the ROP task construction, and re-train BERT itself towards the tailored objective for IR. 
Specifically, we introduce a novel contrastive method, inspired by the divergence-from-randomness idea,  to leverage BERT's self-attention mechanism to sample representative words from the document. By further fine-tuning on downstream ad-hoc retrieval tasks, our method achieves significant improvements over PROP and other baselines, and further pushes forward the SOTA on a variety of ad-hoc retrieval tasks.

\end{abstract}



\maketitle

\section{Introduction}

Recently, we have witnessed the bloom of the `pre-training and fine-tuning' paradigm in the natural language processing (NLP) field   \cite{Brown2020LanguageMA,devlin2018bert,zhang2019pegasus,Raffel2020ExploringTL}, which firstly learns universal language representations from large-scale unlabeled text corpora and then fine-tunes the pre-trained models on downstream tasks.
Pre-trained models have achieved great success in NLP and inspired IR researchers to study their application in IR tasks.
Therefore, researcher investigated to apply pre-trained language models like ELMo \cite{peters-etal-2018-deep}, BERT \cite{devlin2018bert} and T5 \cite{Raffel2020ExploringTL}, for ad-hoc search tasks, and showed that they perform significantly better than previous neural ranking methods \cite{Nogueira2019PassageRW,Dai2019DeeperTU,MacAvaney2019CEDRCE,Nogueira2020DocumentRW,Yang2019SimpleAO}.

Beyond the direct application of existing pre-trained language models designed for NLP, there have been some pioneer studies on the pre-training objectives tailored for IR \cite{Chang2020PretrainingTF,Ma2020PROPPW}. The underlying belief is that the pre-trained model will perform better on a downstream task if its pre-training objective is more similar to that of the downstream task \cite{zhang2019pegasus,Gla2020SpanSP,Ke2020SentiLARESL}.
The latest practice in this direction is the pre-training method with representative words prediction (PROP) \cite{Ma2020PROPPW}, which has reached new SOTA on a variety of ad-hoc retrieval benchmarks \cite{Clarke2009OverviewOT,Clarke2004OverviewOT,Qin2013IntroducingL4} as well as the top performance on the MS MARCO Document Ranking Leaderboard\footnote{https://microsoft.github.io/msmarco/}. 
The key idea of PROP is to construct the \textit{representative words prediction} (ROP) task for pre-training inspired by the query likelihood (QL) model~\cite{Zhai2008StatisticalLM}.
Specifically, word sets are sampled in pairs from documents based on the document language model, and the likelihood of each word set is computed to decide which one is more representative with respect to the document.
The pairwise preference between sampled word sets is then predicted for the ROP task to model the representativeness of words.
PROP also adopts masked language modeling (MLM) to learn good representations for documents.

As we can see, the success of PROP heavily relies on the quality of the ``representative'' words sampled from a document. In their work~\cite{Ma2020PROPPW}, the ROP task is constructed based on the classical multinomial unigram language model with Dirichlet prior smoothing~\cite{Zhai2008StatisticalLM}.
Although the unigram language model is a popular and strong  language model in IR, its limitation is also clear. The term independence assumption adopted by the language model makes it difficult to fully capture the document semantic by ignoring the correlation between words. Consequently, this may affect the ROP sampling process, which is supposed to find representative words with respect to the document semantic.

\begin{table*}[t]

\small
\setlength\tabcolsep{1.5pt}
  \caption{Words sampled from MS MARCO document D360225 using different BERT-based language models. Darker color indicates higher attention. Green color denotes words that are representative for the document while red color denotes the common words. }
  \label{tab:vis}
  \begin{tabular}{p{17.5cm}}
    \hline
    \hline
    \multicolumn{1}{c}{Title: \textbf{pulmonary fibrosis}} \\
    \midrule
     Pulmonary fibrosis Synonyms Interstitial pulmonary fibrosis A chest X-ray demonstrating pulmonary fibrosis believed to be due to amiodarone. Specialty Pulmonology Pulmonary fibrosis (literally ""scarring of the lungs "") is a respiratory disease in which scars are formed in the lung tissues, leading to serious breathing problems. Scar formation, the accumulation of excess fibrous connective tissue (the process called fibrosis ), leads to thickening of the walls, and causes reduced oxygen supply in the blood. As a consequence patients suffer from perpetual shortness of breath. [1]In some patients the specific cause of the disease can be diagnosed, but in others the probable cause cannot be determined, a condition called idiopathic pulmonary fibrosis. There is no known cure for the scars and damage in the lung due to pulmonary fibrosis. [2]Contents [ hide ]1 Signs and symptoms2 Cause3 Pathogenesis4 Diagnosis5 Treatment6 Prognosis7 Epidemiology8 References9 External links Signs and symptoms [ edit]Symptoms of pulmonary fibrosis are mainly: [3]... \\
  
    \midrule    
   \textbf{Vanilla Attention-based Term Distribution}: 
   \colorbox{cgreen!100}{\makebox(35,4){pulmonary}}, \colorbox{cgreen!90}{\makebox(23,4){fibrosis}}, 
   \colorbox{red!80}{\makebox(5,4){in}}, 
   \colorbox{cgreen!70}{\makebox(33,4){interstitial}},  
   \colorbox{red!65}{\makebox(8,4){the}},  
   \colorbox{red!60}{\makebox(5,4){of}}, 
   \colorbox{cgreen!55}{\makebox(22,4){disease}}, 
   \colorbox{red!50}{\makebox(8.5,4){can}}, 
   \colorbox{red!45}{\makebox(8.5,4){and}}, 
   \colorbox{cgreen!40}{\makebox(13,4){lung}}, 
   \colorbox{cgreen!30}{\makebox(13,4){chest}}, 
   \colorbox{red!25}{\makebox(4,4){is}},
  \colorbox{cgreen!20}{\makebox(13,4){cause}}, 
  \colorbox{red!15}{\makebox(6,4){to}},
    \dots
   \\
   
    \midrule
      \textbf{Contrastive Term Distribution}:   
      \colorbox{cgreen!100}{\makebox(23,4){fibrosis}},
      \colorbox{cgreen!90}{\makebox(35,4){pulmonary}},
      \colorbox{cgreen!80}{\makebox(34,4){interstitial}},
      \colorbox{cgreen!70}{\makebox(32,4){idiopathic}},
      \colorbox{cgreen!60}{\makebox(12,4){lung}}, 
      \colorbox{cgreen!50}{\makebox(13,4){chest}}, 
      \colorbox{cgreen!40}{\makebox(22,4){disease}},
      \colorbox{cgreen!35}{\makebox(22,4){diseases}},
      \colorbox{cgreen!30}{\makebox(14,4){cause}}, \colorbox{cgreen!20}{\makebox(22,4){patients}}, 
      \colorbox{cgreen!15}{\makebox(14,4){x-ray}},
      \colorbox{cgreen!10}{\makebox(13,4){scars}},
    \dots
      \\
  \hline
  \hline
\end{tabular}
\end{table*}

To tackle this problem, in this paper, we propose to \textit{Bootstrapped Pre-training with Representative wOrds Prediction} (B-PROP) based on BERT \cite{devlin2018bert} for ad-hoc retrieval. 
As we know, BERT is a contextual language model learned from large-scale unlabeled text corpora. With a stack of multi-head self-attention architecture, each token in BERT accumulates the information from both left and right context to enrich its representation and the special classification token [CLS] is an aggregate of the entire sequence representation, which represents the comprehensive understanding of the input sequence \cite{devlin2018bert}. The BERT representation has shown its superior performance over a large variety of language understanding tasks \cite{Wang2018GLUEAM,Rajpurkar2016SQuAD10,lample-etal-2016-neural}. Therefore, the key idea of B-PROP is to leverage the powerful contextual language model BERT to replace the classical unigram language model for the ROP task construction, and re-train BERT itself towards the tailored objective for IR.

However, it is non-trivial to apply BERT for the ROP task construction. At first thought, an intuitive solution is to directly sample representative words according to BERT's [CLS]-Token attention, which is equivalent to sampling from a term distribution by summing up and re-normalizing the vanilla attention weights over distinct terms. Unfortunately, as shown in Table \ref{tab:vis}, we find that although the vanilla [CLS]-Token attention-based term distribution could generate representative words, it also favors common words, e.g., in, the, of and so on. This is reasonable for BERT since these common words appear frequently and are useful in composing a document, but they are not representative to the document in retrieval. In other words, BERT attempts to encode all the semantic information in a document, so the term distribution obtained from its vanilla attention is a semantic distribution, but not necessarily a representative/informative distribution.

Inspired by the divergence-from-randomness idea \cite{Amati2002ProbabilisticMO}, we propose a novel contrastive method to leverage BERT's attention mechanism to sample representative words from the document. The basic idea of divergence-from-randomness says that \textit{the informativeness of a term could be computed by measuring the divergence between a term distribution produced by a random process and the actual term distribution in a document}. As mentioned above, we can compute the actual term distribution in a document based on BERT's vanilla [CLS]-Token attention. To approximate the term distribution produced by a random process, we take an expectation over all the term distributions in the collection. We then compute the cross-entropy (i.e., the divergence) between the document term distribution and the random term distribution to obtain the \textit{contrastive term distribution}. Sampling from such contrastive term distribution, we can now obtain representative words for a document as shown in Table \ref{tab:vis}. Therefore, in B-PROP, we adopt this contrastive method for sampling to construct the ROP task, and re-train BERT with respect to the PROP learning objective.

We pre-train B-PROP over two kinds of large-scale text corpus respectively following the experimental setting in PROP, i.e., English Wikipedia and MS MARCO Document Collection. 
We then fine-tune B-PROP on $7$ standard ad-hoc retrieval datasets, including 2 large-scale datasets, i.e., MS MARCO document ranking task and TREC 2019 Deep Learning (DL) Track, and 5 small datasets, i.e., Robust04, ClueWeb09-B, Gov2, MQ2007, and MQ2008.
The experimental results showed that B-PROP performs significantly better than other pre-trained models like BERT and PROP, and further pushes forward the SOTA on a variety of ad-hoc retrieval tasks.

\input{background}

\input{model.tex}

\input{exp.tex}

\section{Related Work}


\subsection{Pre-trained Language Models}
Pre-training Transformer-based language models on large-scale unlabeled text corpora and then fine-tuning on downstream tasks have achieved remarkable success in NLP field~\cite{Brown2020LanguageMA,devlin2018bert,Yang2019XLNetGA,zhang2019pegasus,Raffel2020ExploringTL}.
A notable example is BERT~\cite{devlin2018bert},  a bidirectional transformer-based language model that pre-trains with MLM and NSP to learn both contextual text representations and coherence relationship between two pieces of text.
By adding a simple classification layer on top of BERT, BERT could adapt to various classification/regression tasks and outperform many task-specific architectures.
Researchers explored applying these pre-trained models to IR to verify whether they can also help search tasks.
Nogueira \textit{et al.}~\cite{Nogueira2020DocumentRW,Nogueira2019PassageRW,Nogueira2019DocumentEB}, Dai \textit{et al.}~\cite{Dai2019DeeperTU} and MacAvaney \textit{et al.}~\cite{MacAvaney2019CEDRCE} explored to incorporate these pre-trained language models (e.g., BERT) to search tasks. 
A straightforward way is to concatenate the query and the document and then input to the model.
The relevance score is computed using an MLP layer over the representation of the [CLS] token.
They found that BERT could significantly outperform those neural ranking models on ad-hoc retrieval tasks.

\subsection{Pre-training Objectives Tailored for IR}
Beyond the direct application of existing pre-trained language models, there have been some pioneer studies on the pre-training objectives tailored for IR. Besides ICT~\cite{Lee2019LatentRF,Chang2020PretrainingTF}, Chang \textit{et al.}~\cite{Chang2020PretrainingTF} proposed another two pre-training tasks for passage retrieval in open domain question answering, i.e., Body First Selection (BFS) and  Wiki Link Prediction (WLP).
BFS randomly samples a sentence from the first section of a Wikipedia page as a pseudo query, and samples a passage from the rest page as its positive context.
This task aims to capture the semantic relationship outside of a local passage but within a document.
WLP randomly sample a passage from another page in which there is a hyperlink between the two page.
WLP aims to capture inter-page semantic relation.
As we can see, these tasks weakly resemble the relevance relationship between questions and answer passages which may only teach the model to learn the topic-level relevance.
Our experiments showed that pre-training with these tasks for traditional ad-hoc search is not as effective as on the QA.
Recently, Ma \textit{et al.} \cite{Ma2020PROPPW} proposed a novel pre-training objective tailored for ad-hoc retrieval, i.e., representation words prediction (ROP), and achieved new SOTA on a variety of ad-hoc retrieval tasks.
As described in Section~\ref{sec:prop}, PROP might be bounded by the classical language model adopted in the ROP task construction.


\section{Conclusion}

In this paper, we proposed B-PROP to address the problem of PROP where the effectiveness might be bound by the statistical document language model for the ROP task construction.
The key idea of B-PROP is to leverage the powerful contextual language model BERT to replace the unigram language model for the ROP task construction and re-train BERT towards the tailored objective for IR. 
Inspired by the divergence-from-randomness idea, we introduced a contrastive method to measure the divergence between the document term distribution and the random term distribution produced by a random process to compute the informativeness/representativeness of a term.
Therefore, in B-PROP, we adopted this contrastive method for sampling to construct the ROP task, and re-train BERT towards the PROP learning objective.
Experimental results demonstrated the effectiveness of B-PROP on the downstream ad-hoc retrieval tasks. 
In the future work, we would like to adopt ROP objective for the first-stage retrieval in IR.

\section*{Acknowledgments}

This work was supported by Beijing Academy of Artificial Intelligence (BAAI) under Grants No. BAAI2019ZD0306, and funded by the National Natural Science Foundation of China (NSFC) under Grants No. 62006218, 61902381, 61773362, and 61872338, the Youth Innovation Promotion Association CAS under Grants No. 20144310, 2016102, and 2021100, the National Key RD Program of China under Grants No. 2016QY02D0405, the Lenovo-CAS Joint Lab Youth Scientist Project, the K.C.Wong Education Foundation, and the Foundation and Frontier Research Key Program of Chongqing Science and Technology Commission (No. cstc2017jcyjBX0059).

\bibliographystyle{ACM-Reference-Format}
\balance
\bibliography{main}


\end{document}

%% file: background.tex
\section{Background}\label{sec:prop}
\label{background}

Our method adopts the same pre-training task of PROP~\cite{Ma2020PROPPW}, so we would like to first briefly review this method in this section.
The ROP task in PROP is inspired by an old hypothesis underlying the classical statistical language model for IR.
It assumes that the query is a piece of representative text generated from the ``ideal'' document~\cite{Zhai2008StatisticalLM}
Based on this hypothesis, the ROP task is introduced for pre-training for IR as follows.

\textbf{Representative Word Sets Sampling}.  
Word sets are sampled in pairs from documents based on the document language model.
Specifically, PROP samples the length for a pair of word sets from a Poisson distribution first~\cite{azzopardi2007building,arampatzis2008study}.
Then according to the document language model, especially the unigram language model~\cite{Zhai2008StatisticalLM}, they sample pairs of word sets from documents in which two word sets in one pair have the same length.

\textbf{Representative Words Prediction (ROP)}.
The likelihood of each word set is computed based on the QL to decide which one is more representative with respect to the document.
PROP then pre-trains the Transformer by predicting the pairwise preference between pairs of word sets.

Suppose $(s_1,s_2)$ are a pair of word sets sampled from the document $d$,
the likelihood score of the pair of word sets are computed based on QL as follows, 
\begin{equation}\label{likelihood}
score =\prod_{w \in s} P(w|\theta_{d}),
\end{equation}
where $\theta_d$ is the multinomial unigram language model with Dirichlet prior smoothing proposed by \cite{Zhai2008StatisticalLM}.
The ROP task is pre-trained using the hinge loss to predict the pairwise preference as follows
\begin{equation}
\label{ROP}
\mathcal{L}_{ROP} = max(0, 1-P(s_1|d)+P(s_2|d)),
\end{equation}
where the likelihood of each set $P(s|d)$ denotes how representative the word set is to the document. 
It is obtained by applying a MLP over the final hidden state of [CLS] when pre-training.

\textbf{Masked Language Modeling (MLM)}. Similar to BERT, PROP also adopts the MLM~\cite{devlin2018bert} objective as one of its pre-training objectives besides the ROP objective, as it's useful to learn high-quality contextual representations for queries and documents. 
Studies have shown that MLM can learn lexical, syntactic, and semantic information from texts~\cite{Fan2021}.
\begin{equation}\label{mlm}
\begin{aligned}
\mathcal{L}_{MLM} = -\sum_{x_i\in \mathcal{M}} \log p(x_i|\widetilde{x}),
\end{aligned}
\end{equation}
where $\mathcal{M}$ is the masked token set, $x_i$ is one of the masked words from $\mathcal{M}$, and $\widetilde{x}$ is the rest words in $x$.

%% file: model.tex
\section{B-PROP}
In this section, we describe our bootstrapped pre-training method with representative words prediction (B-PROP) for ad-hoc retrieval in detail.

\subsection{Motivation}
The core of the ROP pre-training task mentioned above is to sample representative word sets from a document. As demonstrated in~\cite{Ma2020PROPPW}, a random sampling strategy would lead to much worse performance of the pre-trained model as compared with the document language model-based sampling. This result clearly indicates that the quality of the ``representative'' word sets is key to the success of the pre-training.

Although the unigram language model with Dirichlet prior smoothing has shown its promising results in PROP, its limitation is also clear. The term independence assumption adopted by the language model makes it difficult to fully capture the document semantic by ignoring the correlation between words, which in turn affect the representativeness of the sampled words. To address this problem, we propose to leverage BERT, a powerful contextual language model, to replace the unigram language model for the ROP task construction. The key assumption here is that a better document language model would lead to higher quality of representative words sampling. 


Unfortunately, as shown in the pilot analysis in Table \ref{tab:vis}, directly sampling from the term distribution based on BERT's [CLS]-Token attention may not necessarily produce representative words for a document, since the vanilla attention also favors common words with high frequency, e.g., stop words in IR. The underlying reason is that the contextual language model focuses on encoding as much semantic information in a document as possible, thus becomes less discriminative (i.e., not to differentiate common words which are randomly distributed in the corpus \cite{Amati2002ProbabilisticMO}). Inspired by the divergence-from-randomness idea, we introduce a novel contrastive method to leverage BERT's attention mechanism to sample representative words from a document, and re-train BERT itself towards the PROP learning objective.






\subsection{Contrastive Sampling for ROP with BERT}\label{sec:ctd}



The basic idea of divergence-from-randomness \cite{Amati2002ProbabilisticMO} says that the informativeness/representativeness of a term could be computed by measuring the divergence between a term distribution produced by a random process and the document term distribution. Following this idea, we introduce a contrastive method to leverage BERT’s attention mechanism to sample representative words from a document for the ROP task construction.

In the following, we first show how to compute \textit{the document term distribution} based on BERT’s vanilla [CLS]-Token attention. We then take an expectation over all the term distributions in the collection to approximate \textit{the term distribution produced by a random process}. Finally, we compute the cross-entropy (i.e., the divergence) between the two distributions, i.e., the document term distribution and the random term distribution, to obtain \textit{the contrastive term distribution}. The ROP task is then constructed by sampling pairs of representative word sets from the document.

\begin{algorithm}[t]
\caption{{Representative Word Sets Sampling}}
\label{algo:bprop}
\LinesNumbered
\KwIn{Document Collection $\mathcal{D}$, Vocabulary $V=\{w_k\}_{k=1}^K$, Original BERT model $M$.}
\BlankLine
\For{\textup{each document} $d \in \mathcal{D}$}{
Extract the vanilla [CLS]-Token attention weight $\alpha^t$ from $M$, w.r.t.   Eq. (\ref{alg:dtd1})\\
Compute the document term distribution $P(w_k|\theta_{d})$ based on $\alpha^t$, w.r.t. Eq. (\ref{alg:dtd4})\\
}

\BlankLine
Approximate the random term distribution $P(w_k|\theta_{random})$ based on all  $P(w_k|\theta_{d})$ in $\mathcal{D}$, w.r.t. Eq. (\ref{alg:rtd})\\
\BlankLine

\For{\textup{each document} $d$ in $\mathcal{D}$}{
Compute the contrastive term distribution $P(w_k|\theta_{contrastive})$ based on $P(w_k|\theta_{d})$ and $P(w_k|\theta_{random})$, w.r.t. Eq. (\ref{alg:ctd}) and Eq. (\ref{alg:sctd})\\
Sample the size of the word set $l$, w.r.t. Eq. (\ref{size})\\
Sample a pair of word sets $s_1$ and $s_2$ according to $P(w_k|\theta_{contrastive})$, w.r.t. Eq. (\ref{set})\\
}

\end{algorithm}

\subsubsection{Document Term Distribution}

The document term distribution is computed  based on BERT's vanilla [CLS]-Token attention. 
We first introduce the self-attention mechanism in BERT and then show how to compute the document term distribution.

The multi-head attention in BERT's self-attention layer~\cite{Vaswani2017AttentionIA} allows the model to jointly attend to information from different representation subspaces at different positions. 
Given an input document $d = (x_1,x_2,\dots,x_T)$ with $T$ tokens, which starts with a special classification token ([CLS]), the $i$-th attention head for [CLS] is, 
\begin{equation}
\label{attention}
\small
\nonumber
Attention_i(\textbf{h}_{{\rm [CLS]}}) = \sum_{t=1}^T  softmax(\frac{W_{i}^{q}\textbf{h}_{{\rm [CLS]}} \cdot W_{i}^{k}\textbf{h}_{t}}{\sqrt{d/{h}}})W_{i}^{v}\textbf{h}_{t}, 
\end{equation}  
where $\textbf{h}_{t}$ denotes a $d$ dimensional hidden vector of the $t$-th sequence token. 
$\textbf{h}_{{\rm [CLS]}}$ denotes a $d$ dimensional hidden vector of the [CLS] token. 
$h$ is the number of self-attention heads. 
$W_{i}^{q}$, $W_{i}^{k}$, and $W_{i}^{v}$ are learned matrices of size $d/h \times d$. 
Thus, the the outputs of the $h$ attentions heads for [CLS] are $\{Attention_i(\textbf{h}_{{\rm [CLS]}})\}_{i=1}^h$.

Specifically, we could extract the attention weight $\alpha_i^t$ from the $i$-th attention head for [CLS] in BERT's final layer, i.e., 
\begin{equation}\label{alg:att}
\nonumber
\alpha_i^{t} =	softmax(\frac{W_{i}^{q}\textbf{h}_{{\rm [CLS]}} \cdot W_{i}^{k}\textbf{h}_{t}}{\sqrt{d/{h}}}). 
\end{equation}

Then, we average the [CLS]-Token attention weights across $h$ attention heads,  to obtain the final vanilla [CLS]-Token attention weight $\alpha^t$ of each token, i.e., 
\begin{equation}\label{alg:dtd1}
\alpha{^t} = \frac{1}{h}\sum_{i=1}^h \alpha_i^t.
\end{equation}

Typically, a term may appear multiple times within the same document.  
Based on the vanilla [CLS]-Token attention, for each distinct term $w_k$ in the vocabulary $V=\{w_k\}_{k=1}^K$, we first add up the final vanilla [CLS]-Token attention weights of the same term over different positions in a document $d$, i.e., 
\begin{equation}\label{alg:dtd2}
\nonumber
\hat{\beta}_{w_k} = \sum_{x_t = w_k} \alpha{^t} , x_t \in d.
\end{equation}

The document would be dominated by terms with large attention weights, to alleviate this problem, we leverage the term saturation function used as part of BM25~\cite{Robertson2009ThePR}.
The distinct term attention weight is thus computed by, 
\begin{equation}\label{alg:dtd3}
\nonumber
\beta_{w_k} = \frac{\hat{\beta}_{w_k}}{b+\hat{\beta}_{w_k}},
\end{equation}
where $b$ is a hype-parameter controlling the shape of the saturation curve. Note the distinct term attention weight is set as zero if  the document does not contain the corresponding term.

Then, we normalize the distinct term attention weights of all the terms in the vocabulary $V$, to obtain the document term distribution $P(w_k|\theta_{d})$, 
\begin{equation}\label{alg:dtd4}
P(w_k|\theta_{d}) = \frac{exp(\beta_{w_k})}{\sum_{w_k \in V}{exp(\beta_{w_k})}}.
\end{equation}

\subsubsection{Random Term Distribution}

To approximate the term distribution produced by a random process, we take an expectation over all the term distributions in the document collection $\mathcal{D}$. 
The random term distribution $P(w_k|\theta_{random})$ is computed by,
\begin{equation}\label{alg:rtd}
P(w_k|\theta_{random}) = \mathbb{E}(w_k|\mathcal{D}) = \frac{1}{|\mathcal{D}|} \sum_{d\in \mathcal{D}}P(w_k|\theta_d),
\end{equation}
where $\mathcal{D}$ is the document collection containing $|\mathcal{D}|$ documents.

\subsubsection{Constrastive Sampling for ROP}

Finally, we compute the cross-entropy (i.e., the divergence) between the document term distribution $P(w_k|\theta_{d})$ and the random term distribution $P(w_k|\theta_{random})$ to obtain the contrastive term distribution $P(w_k|\theta_{contrastive})$, i.e.,  
\begin{equation}\label{alg:ctd}
\hat{\gamma}_{w_k} = -P(w_k|\theta_{d})\log_{2}P(w_k|\theta_{random}),  
\end{equation}

\begin{equation}\label{alg:sctd}
P(w_k|\theta_{contrastive}) = \frac{exp(\hat{\gamma}_{w_k})}{\sum_{w_k \in V}{exp(\hat{\gamma}_{w_k}})},
\end{equation}

where the softmax function ensures the contrastive term distribution is valid.
After obtaining the contrastive term distribution, we could sample  pairs of word sets from documents for the ROP task construction.
Algorithm ~\ref{algo:bprop} shows the detailed sampling process of representative words in B-PROP.

First, pairs of word sets are sampled from document $d$ according to the contrastive term distribution. 
Following the procedure in PROP, we sample the length $l$ of a word set from Poisson distribution first \cite{azzopardi2007building,arampatzis2008study},
\begin{equation}\label{size}
l \sim Poisson(\lambda),
\end{equation}
where $\lambda$ is the hyper-parameter. 
Note that two word sets in one pair have the same length for a fair comparison.

Then, a pair of word set $(s_1, s_2)$ are sampled from $d$ according to the contrastive term distribution, i.e.,
\begin{equation}\label{set}
s =  Sample(V), w_k \sim P(w_k|\theta_{contrastive}). 
\end{equation}

\subsection{Re-Training BERT with ROP and MLM}

Based on the above sampled representative word sets, we compute the likelihood score of each word set according to Eq. (\ref{likelihood}).
We then re-train BERT towards the tailored objective for IR, i.e., the ROP objective in Eq. (\ref{ROP}), jointly with the MLM objective in Eq. (\ref{mlm}), as follows, 
\begin{equation}
\label{bprop}
\nonumber
\mathcal{L}_{total}  = \mathcal{L}_{ROP}  + \mathcal{L}_{MLM}.  
\end{equation}

Since we sample representative words from BERT and re-train BERT using the sampled data, such a process is called bootstrapped training and we name the pre-trained model B-PROP.

%% file: exp.tex
\section{EXPERIMENTS}
In this section, we empirically study the effectiveness of B-PROP on 7 standard ad-hoc retrieval datasets.  

\subsection{Experimental Settings}

We first introduce our experimental setting, including datasets,
baseline methods, evaluation methodology, and implementation details.

\subsubsection{Datasets}
Here, we introduce the datasets for pre-training and the downstream tasks.

\begin{itemize}[leftmargin=*]
\item \textbf{Pre-training Corpus}. Following PROP, we also leverage two large public document collections, including Wikipedia and MS MARCO Document Corpus, for pre-training. 
\begin{itemize}[leftmargin=*]
\item \textbf{Wikipedia} is a widely used corpus which contains tens of millions of well-formed Wiki-articles.
\item \textbf{MS MARCO Document Corpus} is another large-scale document collection with about 4 million available Web documents.
\end{itemize} 

By pre-training B-PROP on Wikipedia and MS MARCO Document Corpus respectively, we obtain two types of pre-training models denoted as \textbf{B-PROP$_{\text{Wiki}}$} and \textbf{B-PROP$_{\text{MARCO}}$}. 

\item \textbf{Downstream Tasks}. We fine-tune B-PROP on 7 representative ad-hoc retrieval benchmark collections. The detailed statistics of these downstream datasets are shown in Table~\ref{tab:datasets}.
\begin{itemize}[leftmargin=*]
\item \textbf{Robust04}~\cite{Voorhees2004OverviewOT} is a small news dataset with about 250 queries and 0.5M news articles, where the topics are collected from TREC Robust Track 2004. 
\item \textbf{ClueWeb09-B}~\cite{Clarke2009OverviewOT} is another small dataset with 150 queries collected from TREC Web Tracks 2009-2011 and a large-scale collection with 50M documents.
\item \textbf{Gov2}~\cite{Clarke2004OverviewOT} contains about 150 queries and a large document collection with 25M documents crawled from the .gov domain Web pages, where the topics are collected from TREC Terabyte Tracks 2004-2006. 
\item \textbf{Million Query Track 2007 (MQ2007)} is a learning-to-rank ( LETOR)~\cite{Qin2013IntroducingL4} dataset containing 1692 queries.
It uses the GOV2 corpus as its document collection. 
\item \textbf{Million Query Track 2008 (MQ2008)} is similar to MQ2007 except with a different query set containing 784 queries.
\item \textbf{MS MARCO Document Ranking (MS MARCO)}~\cite{Campos2016MSMA} is a large-scale benchmark dataset for web document retrieval, with about 0.37 million training queries.
\item \textbf{TREC 2019 Deep Learning Track (TREC DL)}~\cite{Craswell2020OverviewOT} replaces the test set in MS MARCO with a novel set produced by TREC with more comprehensive labeling. 
\end{itemize}


\end{itemize}

\begin{table}[t]
\renewcommand{\arraystretch}{1.1}
\setlength\tabcolsep{8pt}
  \caption{Statistics on 7 ad-hoc retrieval datasets}
  \label{tab:datasets}
  \begin{tabular}{ccccc}
    \toprule
    \toprule
    Dataset & Type & \#Queries & \#Documents \\
    \midrule
    MS MARCO & web pages & 0.37M & 3.2M	 \\
    TREC DL & web pages & 0.37M & 3.2M \\
    \midrule
    Robust04 & news & 250 & 0.5M\\ 
    ClueWeb09-B & web pages & 150 & 50M\\ 
    Gov2 & .gov pages & 150 & 25M \\ 
    MQ2007 & .gov pages & 1,692 & 25M \\ 
    MQ2008 & .gov pages & 784 & 25M \\ 

  \bottomrule
  \bottomrule
\end{tabular}
\end{table}

\subsubsection{Baselines}
We consider three groups of baseline models for performance comparison, including traditional retrieve models, popular neural IR models, and advanced pre-trained models.

\begin{itemize}[leftmargin=*]
\item \textbf{Traditional Retrieval Models} include:
\begin{itemize}[leftmargin=*]
\item \textbf{Query likelihood model (QL)} \cite{Zhai2008StatisticalLM} is a representative method of classical language modeling approach.
\item \textbf{BM25} \cite{Robertson2009ThePR} is a strong probabilistic retrieval model which has shown to be very strong across different IR tasks.
\end{itemize}

\item \textbf{Neural IR Models} include:
\begin{itemize}[leftmargin=*]
\item \textbf{DRMM} \cite{Guo2016ADR} is a deep relevance matching model which employs a joint deep architecture at the query term level for relevance matching. 
\item \textbf{Conv-KNRM} \cite{Dai2018ConvolutionalNN} is a popular Convolutional Kernel-based Neural Ranking Model which  models n-gram soft matches for ad-hoc retrieval.
\end{itemize}

\item \textbf{Pre-trained Models} include:
\begin{itemize}[leftmargin=*]
\item \textbf{BERT} \cite{devlin2018bert} is designed to pre-train deep bidirectional representations from large-scale unlabeled text with MLM and Next Sentence Prediction (NSP). 

\item \textbf{Transformer$_{ICT}$} pre-trains with Inverse Cloze Task (ICT)~\cite{Chang2020PretrainingTF} which is designed for passage retrieval in open domain question answering. 
ICT randomly samples a sentence from a passage and treats the rest sentences as its positive context.
We pre-train the Transformer model on Wikipedia with both ICT and MLM objectives.

\item \textbf{PROP} is a recently proposed pre-training model tailored for IR. There are two types of PROP models, i.e., PROP$_{Wiki}$ and PROP$_{MARCO}$, which is pre-trained on Wikipedia and MS MARCO Document Ranking dataset respectively.

\end{itemize}
\end{itemize}

\begin{table*}[t]
\renewcommand{\arraystretch}{1.1}
\setlength\tabcolsep{2pt}
 \caption{Performance Comparisons between B-PROP and the baselines on 5 small datasets.  Two-tailed t-tests demonstrate the improvements of B-PROP to the best baseline PROP are statistically significant ( $\ast$ indicates $p \le 0.05$).}
  \label{tab:main_res1}
  \begin{tabular}{cccccccccccccccc}
    \toprule
    \toprule
    \multirow{2}{*}{Model Type} &\multirow{2}{*}{Model Name} & \multicolumn{2}{c}{Robust04} && \multicolumn{2}{c}{ClueWeb09-B} && \multicolumn{2}{c}{Gov2}  && \multicolumn{2}{c}{MQ2007} && \multicolumn{2}{c}{MQ2008}\\
    \cline{3-4} \cline{6-7} \cline{9-10}  \cline{12-13}  \cline{15-16}  
  & & nDCG@20 & P@20 & &nDCG@20 & P@20 && nDCG@20 & P@20 && nDCG@10 & P@10&& nDCG@10 & P@10  \\
    \midrule 
     \multirow{2}{*}{\shortstack{Traditional \\ Retrieval Models}} & QL & 0.413 & 0.367 && 0.225 &0.326 & & 
     0.409 &0.510&&0.423&0.371&&0.223&0.241\\
     & BM25 & 0.412 &0.363& &0.230 & 0.334& &0.421 &0.523&&0.414&0.366&&0.220&0.245\\
    \midrule
     \multirow{2}{*}{\shortstack{Neural IR \\Models}} & DRMM & 0.425 & 0.371 && 0.246 &0.349 && 
     0.457 & 0.545 && 0.441 & 0.382 && 0.221 & 0.248\\
    & Conv-KNRM & 0.414 & 0.360 && 0.238 & 0.336 && 0.462 &0.552 && 0.431 & 0.377 && 0.215 & 0.239\\
    \midrule
      \multirow{4}{*}{\shortstack{Pre-trained \\ Models}} & BERT& 0.459  & 0.389 && 0.295 & 0.367 && 0.495 & 0.586 && 0.506 & 0.419 &&0.247 &0.256 \\
      & Transformer$_{ICT}$ & 0.460 & 0.388 && 0.298 & 0.369 && 0.499&0.587 && 0.508& 0.420 && 0.245& 0.256\\
      & PROP$_{Wiki}$ & 0.502 & 0.421 && 0.316 & 0.384 && 0.519 & 0.593 && 0.523 & 0.432 && 0.262 &0.267 \\
      & PROP$_{MARCO}$ & 0.484 & 0.408 && 0.329 & 0.391 && 0.525 & 0.594 && 0.522 & 0.430 && 0.266 &0.269 \\
    \midrule
    \multirow{2}{*}{Our Approach} 
    & B-PROP$_{Wiki}$ & \textbf{0.519}$^{\ast}$  & \textbf{0.430}$^{\ast}$ && 0.331 & 0.393 && 0.534$^{\ast}$& 0.599$^{\ast}$ && \textbf{0.529}$^{\ast}$&0.436$^{\ast}$ &&0.271$^{\ast}$ & 0.273 \\
    & B-PROP$_{MARCO}$ & 0.510$^{\ast}$ & 0.429$^{\ast}$ & & \textbf{0.353}$^{\ast}$ & \textbf{0.407}$^{\ast}$ &
    & \textbf{0.552}$^{\ast}$ & \textbf{0.606}$^{\ast}$ && \textbf{0.529}$^{\ast}$ & \textbf{0.439}$^{\ast}$ && \textbf{0.273}$^{\ast}$ & \textbf{0.275}$^{\ast}$ \\
    \bottomrule
    \bottomrule
  \end{tabular}
\end{table*}

\subsubsection{Evaluation Methodology}

For 5 small datasets, we conduct 5-fold cross-validation following previous words~\cite{Fan2018ModelingDR,Guo2016ADR}. 
We take the topic ``titles'' as queries for 5 small datasets and queries are randomly divided into 5 folds.
For each run, we train our model on 3 folds and use another one fold as a validation set to tune parameters.
We take the best checkpoint on the validation set to evaluate the last fold.
This process is repeated 5 times, once for each fold. The final result is obtained by averaging the performance on each tested fold.
Following~\cite{Guo2016ADR,Dai2019DeeperTU,MacAvaney2019CEDRCE, Ma2020PROPPW}, we report the precision at rank 20 (P@20) and normalized discounted cumulative gain at rank 20 (nDCG@20) for Robust04, ClueWeb09-B and Gov2. 
Following~\cite{Fan2018ModelingDR,Pang2017DeepRankAN}, we report the P@10 and NDCG@10 for MQ2007 and MQ2008. 

For two large-scale ad-hoc retrieval collections, the top 100 documents are compared using the Mean Reciprocal Rank at 100 (MRR@100) for MS MARCO and nDCG@10 for TREC DL, which is  suggested in the official instructions. 
We also evaluate MRR@10 and nDCG@100 for MS MARCO and TREC DL, respectively. 
For MS MARCO, we report the performance results on the dev set since  the MS MARCO Document Ranking leaderboard limits the frequency of the submission. 
For TREC DL, we report results on the test set.

\subsubsection{Implementation Details}

We describe the implementation details of B-PROP and baselines, pre-training procedures, and fine-tuning procedures.

\begin{itemize}[leftmargin=*]

\item \textbf{Model Implementations}. 

The pre-trained models (i.e., BERT, Transformer$_{ICT}$, PROP and B-PROP) are implemented based on the Transformers toolkit~\footnote{https://github.com/huggingface/transformers}. 
For Transformer$_{ICT}$ and PROP, parameters are initialized from the BERT's checkpoint released by Google\footnote{https://github.com/google-research/bert}. 
For B-PROP, since BERT applies WordPiece tokenization which could split an entire word into several subwords, we take the [CLS]-Token attention weight of the first subword as that of the entire word. 
The parameter $b$ in the saturation function is set as $0.01$.

\item \textbf{Pre-training Procedures}. For MLM, we randomly mask 15\% words in the input sequence following BERT.
The details of MLM refer to the original paper~\cite{devlin2018bert}.
Note that the sampling is based on the tokenized words and we do not mask the sampled word set for a concatenated text sequence.
The input sequence are truncated to a maximum of 512 for all pre-trained models. 
We use Adam optimizer \cite{Kingma2015AdamAM} with a linear warm-up technique over the first 10\% steps to learn B-PROP's parameters. 
We sample 5 pairs of word sets for each document according to the contrastive term distribution in the document. 
We pre-train B-PROP with a batch size of 128 and a learning rate of 5e-5. 
The pre-training procedures usually take about 6 days on up to 4 Nvidia Telsa V100-32GB GPUs. 

\item \textbf{Fine-tuning Procedures}. For 5 small datasets, we only test the models on the re-ranking stage as with previous works \cite{Guo2016ADR,Pang2017DeepRankAN,Fan2018ModelingDR}.
We retrieve the top 200 candidate documents for each query with the BM25 model on the Anserini toolkit \footnote{https://github.com/castorini/anserini}.
Then all the pre-trained models are fine-tuned on these retrieved documents. 
Following PROP, we also fine-tune the model with a batch size in the range of [16,32] and the learning rate in the range of [$1e-5$,$2e-5$].

For 2 large-scale datasets, we follow the official setting \cite{Craswell2020OverviewOT} to take both \textit{rerank} and \textit{fullrank} subtasks for the evaluation. For the rerank subtask, we run all models by re-ranking the top 100 candidate documents provided by the official MS MARCO and TREC team. For the fullrank subtask, we firstly take the doc2query~\cite{Nogueira2019DocumentEB} model to expand all documents with generated pseudo questions as it has been proved to be able to improve the recall performance \cite{Lin2020PretrainedTF}. Then, we retrieve the top 100 ranked documents using BM25 on the expanded document corpus for evaluation.
It is worth to note that the \textit{fullrank} here is defined as taking the user-defined retrieval system for the phrase 1 retrieval, rather than to implement full end-to-end retrieval\cite{Craswell2020OverviewOT}.  
For all the pre-trained models, we set the learning rate as $1e-5$, the batch size as 144, and the number of fine-tuning epochs as 10.


\end{itemize}

\begin{table*}[t]
  \renewcommand{\arraystretch}{1.1}
  \setlength\tabcolsep{2.5pt} 
  \caption{Comparisons between B-PROP and the baselines on 2 large-scale datasets. Two-tailed t-tests demonstrate the improvements of B-PROP to the best baseline PROP are statistically significant ( $\ast$ indicates $p \le 0.05$).}
  \label{tab:main_res2}
  \begin{tabular}{cccccccccc}
  \toprule
    \toprule
    \multirow{4}{*}{Model Type} & \multirow{4}{*}{Model Name} & \multicolumn{4}{c}{MS MARCO} & \multicolumn{4}{c}{TREC DL} \\ 
    \cmidrule(lr){3-6} \cmidrule(lr){7-10} 
    & & \multicolumn{2}{c}{rerank} & \multicolumn{2}{c}{fullrank} & \multicolumn{2}{c}{rerank} & \multicolumn{2}{c}{fullrank}\\ 
    \cmidrule(lr){3-4} \cmidrule(lr){5-6} \cmidrule(lr){7-8} \cmidrule(lr){9-10} 
     &  & MRR@10 & MRR@100 &  MRR@10 & MRR@100 & nDCG@10  & nDCG@100 & nDCG@10  & nDCG@100  \\ 
    \midrule
    \multirow{2}{*}{\shortstack{Traditional \\Retrieval Models}} & QL & - & - & 0.287 & 0.300 & - & -& 0.600 & 0.559  \\
    & BM25 & - & - & 0.315 & 0.326 & - & - & 0.592 & 0.552   \\
    \midrule
    \multirow{2}{*}{\shortstack{Neural IR \\Models}} & DRMM & 0.137 & 0.152 & 0.164 &  0.197 &  0.249 & 0.390 & 0.301 & 0.422 \\
    & Conv-KNRM & 0.155 & 0.179 & 0.183 & 0.225 & 0.311 & 0.476 & 0.360 & 0.456 \\
    \midrule
    \multirow{4}{*}{\shortstack{Pre-trained \\ Models}} &BERT  & 0.391 & 0.397 & 0.410 & 0.418 & 0.642  & 0.519 & 0.657 & 0.567 \\
    & Transformer$_{ICT}$ &  0.394 & 0.399 & 0.411 & 0.423 & 0.639 & 0.521 & 0.658 & 0.569  \\
    & PROP$_{Wiki}$ & 0.401 & 0.405 & 0.419 & 0.427 & 0.654 & 0.533 & 0.662 & 0.572  \\
    & PROP$_{MARCO}$ & 0.410 & 0.415 & 0.426 & 0.435 & 0.668 & 0.547 & 0.676 &  0.573 \\
    \midrule
    \multirow{2}{*}{Our Approach} & B-PROP$_{Wiki}$  & 0.415$^{\ast}$ & 0.419$^{\ast}$ & 0.428 & 0.439$^{\ast}$ & 0.670 & 0.552$^{\ast}$ & 0.679 & 0.581$^{\ast}$  \\
    & B-PROP$_{MARCO}$  & \textbf{0.419}$^{\ast}$ & \textbf{0.423}$^{\ast}$ & \textbf{0.437}$^{\ast}$ & \textbf{0.441}$^{\ast}$ & \textbf{0.675}$^{\ast}$ & \textbf{0.558}$^{\ast}$ & \textbf{0.694}$^{\ast}$ & \textbf{0.590}$^{\ast}$  \\
    \bottomrule
    \bottomrule
  \end{tabular}
\end{table*}

\subsection{Baseline Comparison}\label{exp:mai_result}

In this section, we first look at the performance comparisons between \textit{B-PROP}  and the baselines on the 5 small datasets. 
As shown in Table~\ref{tab:main_res1}, we can observe that: 
(1) Compared with traditional retrieval models and neural IR models, the pre-trained model \textit{BERT}  can improve the performance significantly across all the tasks especially on datasets with very few training instances.
For example, the relative improvements of \textit{BERT} over BM25 and DRMM are about  $28\%$ and $20\%$ respectively in terms of nDCG@20 on ClueWeb09-B. 
These results show that pre-training on large-scale text corpus could learn useful text representations to benefit ranking tasks.
(2) \textit{Transformer$_{ICT}$} obtains similar performance with the \textit{BERT} model on all 5 datasets, indicating that the ICT objective designed for passage retrieval in QA scenario cannot bring additional benefits to ad-hoc document retrieval as compared with BERT. 
(3) \textit{PROP} achieves the best performance among all the baseline methods, demonstrating that the pre-trained model will perform better on a downstream task if its pre-training objective is more similar to that of the downstream task.
For example,  the relative improvements of \textit{PROP} over \textit{BERT} and \textit{Transformer$_{ICT}$} are about $12\%$ and $10\%$ respectively in terms of nDCG@20 on ClueWeb09-B.

When we look at \textit{B-PROP}, we find that: 
(1) Our \textit{B-PROP} model outperforms the best baseline \textit{PROP} significantly.  
The better results of our models over \textit{PROP} demonstrate the effectiveness of leveraging the powerful contextual language model \textit{BERT} to replace the classical unigram language model for the ROP task construction. 
(2) Among the two variants of our model,  \textit{B-PROP$_{Wiki}$} performs better than  \textit{B-PROP$_{MARCO}$} on Robust04, while  \textit{B-PROP$_{MARCO}$} performs better than \textit{B-PROP$_{Wiki}$} on ClueWeb09-B and Gov2. 
This observation is consistent in PROP in which the pre-trained model would perform better if the pre-trained corpus and datasets of IR tasks are in a similar domain.
Other studies show that the scale of the pre-training corpus also affects the fine-tuning performance.

Then, we compare the performance between B-PROP and the baselines on 2 large-scale datasets. 
As shown in Table~\ref{tab:main_res2}, we can see that:
(1) The performance trend of different types of models on these two large-scale datasets is quite consistent with that on the previous small datasets. Note that neural IR models perform poorly on these 2 datasets indicating that they are under-fitting due to the small model size. 
(2) The improvements of \textit{B-PROP$_{Wiki}$} over \textit{BERT} and \textit{PROP$_{Wiki}$} are around $5\%$ and $3\%$ respectively in terms of MRR@100 on MS MARCO under both the rerank and fullrank settings. 
The results indicate that B-PROP can benefit downstream ad-hoc retrieval tasks not only on small-sized training data, but also on large-scale training data. 
(3) Among the two variants of our models, \textit{B-PROP$_{MARCO}$} performs better than \textit{B-PROP$_{Wiki}$} in terms of all the evaluation metrics on all the two datasets. 
This further confirms the importance of the consistency between the fine-tuning document corpus and the pre-training document corpus.

Finally, by comparing the performance of \textit{B-PROP} on all datasets, we find that the improvements of \textit{B-PROP} over the baselines on small-sized datasets are significantly higher than that on large-scale datasets. 
For example, the improvement of \textit{B-PROP$_{MARCO}$} over \textit{BERT} is about $20\%$ on ClueWeb09-B in terms of nDCG@20, while the improvement of \textit{B-PROP$_{MARCO}$} over \textit{BERT} is about $6\%$ in terms of MRR@100 on MS MARCO under the \textit{fullrank} subtask. 
The reason might be that the large-scale labeled data is sufficient to train an accurate enough model, and thus they may gain little benefit from the pre-trained models. 
However, it is often time-consuming and expensive to annotate high-quality data for IR.
Therefore, it is of great value to explore the pre-training objectives tailored for ad-hoc retrieval since most real-world IR applications often have limited labeled data.

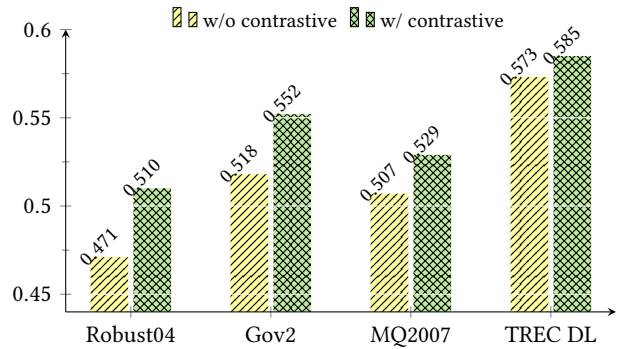
\begin{figure}[t]
\centering 
\begin{tikzpicture}
 \begin{axis}[
  ybar,
  axis on top,
        height=.3\textwidth,
        width=0.5\textwidth,
        bar width=0.5cm,
        ytick={0.45 , 0.5, 0.55, 0.6},
        ymajorgrids, tick align=inside,
        major grid style={draw=white},
        minor y tick num={1},
        enlarge y limits={value=.1,upper},
        ymin=0.44, ymax=0.6,
        axis lines=left,
        enlarge x limits=0.155,
        legend style={
            at={(0.5,1.1)},
            font=\small,
            anchor=north,
            draw=none,
            legend columns=-1,
            /tikz/every even column/.append style={column sep=0.1cm}
        },
        ylabel style={
            anchor=south,
            at={(ticklabel* cs:0.5)},
            yshift=-15pt
        },
        symbolic x coords={Robust04, Gov2, MQ2007, TREC DL}, xtick=data,
        nodes near coords={\pgfmathprintnumber[fixed zerofill,precision=3]{\pgfplotspointmeta}},
      every node near coord/.append style={anchor=south,rotate=45, font=\fontsize{8pt}{6pt}\selectfont},
 ]
 \addplot [draw=none, fill=c1, postaction={pattern=north east lines}] coordinates {
      (Robust04, 0.471)
      (Gov2, 0.518)
      (MQ2007, 0.507)
      (TREC DL, 0.573)
      };
  \addplot [draw=none,fill=c2, postaction={pattern= crosshatch}] coordinates {
      (Robust04,0.51)
      (Gov2, 0.552)
      (MQ2007, 0.529)
      (TREC DL, 0.585)
      };
      \legend{w/o contrastive, w/ contrastive}
 \end{axis}

\end{tikzpicture}
\caption{Comparison of pre-training \textit{B-PROP$_{MACRO}$} with and without contrastive methods. We report nDCG@20 for Robust04 and Gov2, and nDCG@10 for MQ2007 and TREC DL.}
\label{fig:contrastive}
\end{figure}

\subsection{Impact of Contrastive Method}

Since we propose a contrastive method to sample representative words for the ROP task construction, we here study the effect of the contrastive method. 
Specifically, we conduct experiments to compare the performance of \textit{B-PROP$_{MACRO}$} by directly sampling representative words from the document term distribution (i.e., Eq. (\ref{alg:dtd4})) and the contrastive term distribution (i.e., Eq. (\ref{alg:ctd})) in the ROP task.  
Due to the limited space, we only report the nDCG@20 results on Robust04 and Gov2, and the nDCG@10 results on MQ2007 and TREC DL. 
As shown in Figure~\ref{fig:contrastive}, we can find that the performance of \textit{B-PROP$_{MACRO}$} by sampling representative words from the contrastive term distribution is better than that from the document term distribution.  
As discussed above, we have found that the document term distribution based on BERT's vanilla [CLS]-Token attention favors  common words, which may not necessarily produce representative words. 
These results further confirm the effectiveness of the contrastive method by computing cross-entropy between the document term distribution and the random term distribution to obtain a representative/informative term distribution.

\subsection{Impact of Saturation Function} 

A saturation function (i.e., Eq. \ref{alg:dtd3}) borrowed from BM25 is applied to help obtain a good document term distribution.    
The underlying belief is any one term’s contribution to the document score cannot exceed a saturation value, no matter how large it has the vanilla [CLS]-Token attention weight.
Here, we study how such saturation function affects the model performance. 
Specifically, we pre-train \textit{B-PROP$_{MACRO}$} on the \textit{fullrank} subtask with different values of $b$ to control the shape of the saturation curve for analysis. 
As shown in Table~\ref{tab:saturation}, we can find: 
For performance with $b$ set as 0.05, 0.01, and 0.001, we can see that this parameter has a strong affect on the model performance. Actually, it would lead to the over smooth or less smooth on the saturation function by leveraging either a large $b$ or a small $b$. Our results shows that the best performance can be achieved when $b$ is set to $0.01$. 
A better strategy is to set it as the most frequent value according to the frequence distribution of attention weights.
By varying $b$ in the range of 0.05, 0.01, and 0.001, the performance could also be affected by the value of $b$, showing the importance of this hyper-parameter.

\begin{table}[t]
  \caption{Comparison of \textit{B-PROP$_{MACRO}$} pre-training with and without applying the saturation function (SAT), and using SAT with different $b$. Two-tailed t-tests demonstrate the \textit{degradation} with respect to (w/SAT, k=0.01) method are statistically significant ( $-$ indicates $p \le 0.05$).}
  \label{tab:saturation}
  \begin{tabular}{lcccc}
    \toprule
    \toprule
   \multirow{2}{*}{} &  Robust04 & ClueWeb09-B & MS MARCO  \\
   & (nDCG@20) & (nDCG@20) & (MRR@100)  \\
    \midrule
    w/o SAT & 0.469$^-$ & 0.308$^-$  & 0.424$^-$  \\
    w/ SAT, b=0.05 & 0.486$^-$  & 0.319$^-$  & 0.432$^-$ \\
    w/ SAT, b=0.01 & \textbf{0.510} & \textbf{0.353} & \textbf{0.441} \\
    w/ SAT, b=0.001 & 0.475$^-$  & 0.313$^-$  & 0.426$^-$  \\
  \bottomrule
  \bottomrule
\end{tabular}
\end{table}

\subsection{Impact of Pre-training Steps}\label{sec:ckpt}
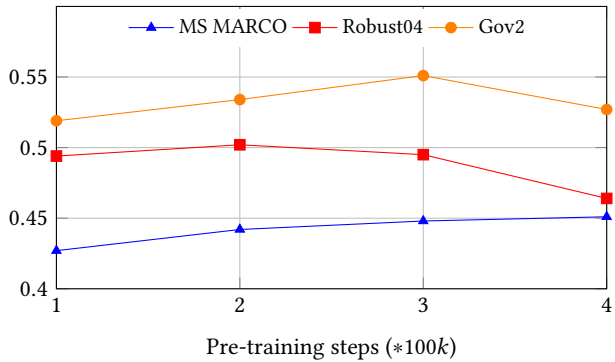
\begin{figure}[t]
\centering 
\begin{tikzpicture}
\begin{axis}[
height=0.3\textwidth,
width=0.5\textwidth,
ytick={0.4, 0.45, 0.5, 0.55},
xtick={1,2,3,4},
ymajorgrids, tick align=inside,
xmajorgrids, tick align=inside,
major grid style,
xmin=1, xmax=4,
ymin=0.4, ymax=0.6,
xlabel=Pre-training steps ($*100k$),
legend columns=3,
legend style={draw=none, at={(0.115,0.925)},anchor=west, nodes={scale=0.9, transform shape}} %
] 
\addplot [mark=triangle*, blue] 
table
{
 X Y
 1 0.427
 2 0.442
 3 0.448
 4 0.451
};
\addplot [mark=square*, red] 
table 
{  
 X Y
 1 0.494
 2 0.502
 3 0.495
 4 0.464
};
\addplot [mark=*, orange] 
table 
{  
 X Y
 1 0.519
 2 0.534
 3 0.551
 4 0.527
};
\addlegendentry{MS MARCO} 
\addlegendentry{Robust04}
\addlegendentry{Gov2}
\end{axis}
\end{tikzpicture}
\caption{The performance of \textit{B-PROP$_{MACRO}$} on different downstream tasks over pre-training steps.}
\label{fig:ckpt}
\end{figure}

We find that pre-training steps may effect the performance of different downstream tasks.
As shown in Figure~\ref{fig:ckpt}, we report MRR@100 for MS MARCO and nDCG@20 for Robust04 and Gov2.
We can see that:
(1) The performance on MS MARCO in terms of MRR@100 under the \textit{fullrank} setting increases over the pre-training steps. 
Moreover, with the same pre-training steps on MS MARCO Document Corpus, \textit{B-RPOP} could achieve \textbf{0.449} in terms of MRR@100 which outperforms \textit{PROP\_step400K\ base and long\ query + doc2query\ top100 (single)} on the MS MARCO Document Ranking leaderboard \footnote{https://microsoft.github.io/msmarco/}.
(2) The performances on Robust04 and Gov2 in terms of nDCG@20  firstly increase with a small number of pre-training steps, and then decreases with more pre-training steps. For example, the best performance on Robust04 is achieved when pre-training $200k$ steps while the best performance on Gov2 is achieved when pre-training $300k$ steps. 
All these results indicate that pre-training with more steps could lead to better fine-tuning performance on the downstream tasks with the similar/same domain, and it may also result in overfitting on the source domain.

\begin{figure*}[t]
	\centering
		\includegraphics[scale=0.48]{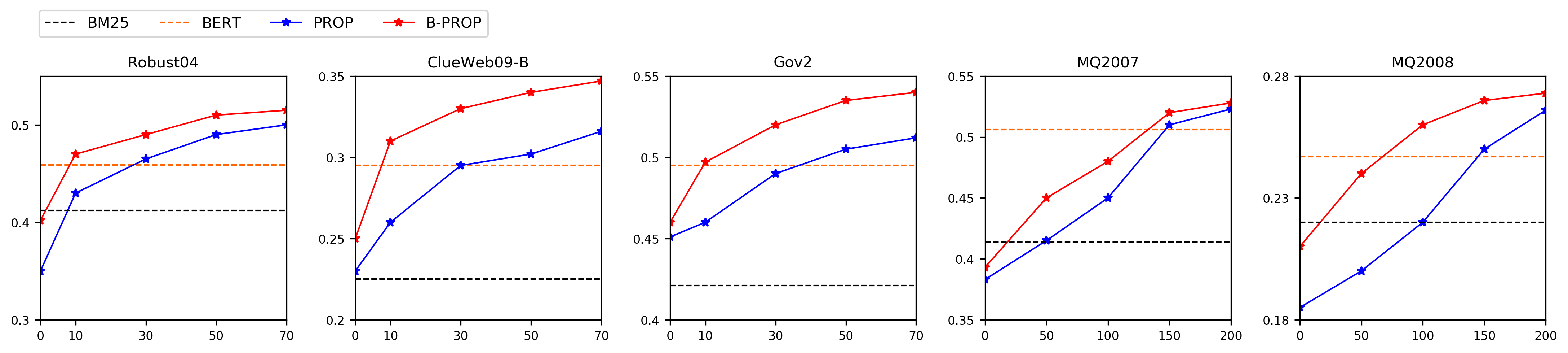}
		  \caption{Performance comparison under zero and low resource setting. The x-axis indicates the number of training queries.}
  \label{fig:fs}
\end{figure*}

\begin{table}[t]
  \caption{Comparison of transfer learning from supervised data and pre-training. Two-tailed t-tests demonstrate the \textit{degradation} with respect to B-PROP$_{MARCO}$ are statistically significant ( $-$ indicates $p \le 0.05$)}
  \label{tab:transfer}
  \begin{tabular}{ccccc}
    \toprule
    \toprule
   \multirow{2}{*}{Model} &  \multicolumn{2}{c}{Robust04} & \multicolumn{2}{c}{Gov2}  \\
      \cmidrule(lr){2-3} \cmidrule(lr){4-5} 
   & nDCG@20 & P@20 & nDCG@20 & P@20  \\
    \midrule
    BERT & 0.459$^-$ & 0.389$^-$ & 0.495$^-$ & 0.586$^-$ \\
    BERT$_{MS MARCO}$ & 0.462$^-$ & 0.391$^-$ & 0.512$^-$ & 0.590$^-$ \\
    B-PROP$_{MARCO}$  & \textbf{0.510} & \textbf{0.429} & \textbf{0.552} & \textbf{0.606} \\
  \bottomrule
  \bottomrule
\end{tabular}
\end{table}

\subsection{Comparison of Transfer Learning from Supervised Data}

Here, we compare pre-training on unlabeled corpus (i.e., documents) with transfer learning from supervised data (i.e., query-document pairs).
We first train BERT on MS MARCO with labeled query-document pairs and then fine-tune this model on the Robust04 and Gov2 datasets. 
We denote this model as \textit{BERT$_{MS MARCO}$}. 
For \textit{B-PROP$_{MARCO}$} and \textit{BERT}, we directly fine-tune the model on the Robust04 and Gov2 datasets.

As shown in Table~\ref{tab:transfer}, we can see that: 
(1) \textit{BERT$_{MS MARCO}$} improves the performance over \textit{BERT} with only a small margin. 
The reason might be that the learning objective of \textit{BERT$_{MS MARCO}$} on the MS MARCO is the same as the final ranking objective on the Robust04 and Gov2. 
Such transfer learning methods only work if the initial and target problems are similar enough.
While the knowledge in the MS MARCO doesn't seem to be effectively transferred to the Robust04 and Gov2 due to the difference of the domain, query type, and matching pattern. 
(2) By pre-training towards the MLM and ROP objective, \textit{B-PROP$_{MARCO}$} achieves the best performance, which again demonstrate its effectiveness. 
This indicates that our pre-training is a more effective and generalizable method which can benefit a variety of downstream tasks.

\subsection{Zero and Low Resource Setting}

Data labeling in IR is super expensive as there is always a very large document collection ranging from millions to billions.
So in this experiment, we aim to show whether B-PROP can perform well with limited data.
We fine-tune B-PROP with a limited number of queries on 5 small ad-hoc retrieval datasets to verify its effectiveness.
For 5 small datasets, we fine-tune \textit{B-PROP} with the batch size as 8, and choose the best learning rate in the range of  [1e-5, 2e-5]  on the validation set without warm-up.

As shown in Figure~\ref{fig:fs}, we report the nDCG@20 results for the Robust04, ClueWeb09-B and Gov2, and the nDCG@10 results for the MQ2007 and MQ2008.
We can see that: 
(1) \textit{B-PROP} outperforms \textit{PROP} significantly on all the datasets using the same number of limited supervised data. 
(2) Fine-tuning \textit{B-PROP} with a limited number of queries can perform on a par with \textit{BERT} fine-tuned on the full supervised datasets.
For example, \textit{B-PROP} fine-tuned with 10 queries could outperform \textit{BERT} fine-tuned on the full supervised datasets on Robust04, ClueWeb09-B and Gov2. 
(3) B-PROP also achieves surprising results in a zero-shot setting.
For instance, \textit{B-PROP} could outperforms BM25 on ClueWeb09-B and Gov2.

%% file: main.bbl

\begin{thebibliography}{39}


\ifx \showCODEN    \undefined \def \showCODEN     #1{\unskip}     \fi
\ifx \showDOI      \undefined \def \showDOI       #1{#1}\fi
\ifx \showISBNx    \undefined \def \showISBNx     #1{\unskip}     \fi
\ifx \showISBNxiii \undefined \def \showISBNxiii  #1{\unskip}     \fi
\ifx \showISSN     \undefined \def \showISSN      #1{\unskip}     \fi
\ifx \showLCCN     \undefined \def \showLCCN      #1{\unskip}     \fi
\ifx \shownote     \undefined \def \shownote      #1{#1}          \fi
\ifx \showarticletitle \undefined \def \showarticletitle #1{#1}   \fi
\ifx \showURL      \undefined \def \showURL       {\relax}        \fi
\providecommand\bibfield[2]{#2}
\providecommand\bibinfo[2]{#2}
\providecommand\natexlab[1]{#1}
\providecommand\showeprint[2][]{arXiv:#2}

\bibitem[\protect\citeauthoryear{Amati and Rijsbergen}{Amati and
  Rijsbergen}{2002}]%
        {Amati2002ProbabilisticMO}
\bibfield{author}{\bibinfo{person}{G. Amati} {and} \bibinfo{person}{C.
  Rijsbergen}.} \bibinfo{year}{2002}\natexlab{}.
\newblock \showarticletitle{Probabilistic models of information retrieval based
  on measuring the divergence from randomness}.
\newblock \bibinfo{journal}{\emph{ACM Trans. Inf. Syst.}}
  (\bibinfo{year}{2002}), \bibinfo{pages}{357--389}.
\newblock


\bibitem[\protect\citeauthoryear{Arampatzis and Kamps}{Arampatzis and
  Kamps}{2008}]%
        {arampatzis2008study}
\bibfield{author}{\bibinfo{person}{Avi Arampatzis} {and} \bibinfo{person}{Jaap
  Kamps}.} \bibinfo{year}{2008}\natexlab{}.
\newblock \showarticletitle{A Study of Query Length}. In
  \bibinfo{booktitle}{\emph{Proceedings of the 31th Annual International ACM
  SIGIR Conference on Research and Development in Information Retrieval}}.
  \bibinfo{publisher}{ACM}, \bibinfo{address}{New York, NY, USA},
  \bibinfo{pages}{811–812}.
\newblock


\bibitem[\protect\citeauthoryear{Azzopardi, de~Rijke, and Balog}{Azzopardi
  et~al\mbox{.}}{2007}]%
        {azzopardi2007building}
\bibfield{author}{\bibinfo{person}{Leif Azzopardi}, \bibinfo{person}{Maarten de
  Rijke}, {and} \bibinfo{person}{Krisztian Balog}.}
  \bibinfo{year}{2007}\natexlab{}.
\newblock \showarticletitle{Building Simulated Queries for Known-Item Topics:
  An Analysis Using Six European Languages}. In
  \bibinfo{booktitle}{\emph{Proceedings of the 30th Annual International ACM
  SIGIR Conference on Research and Development in Information Retrieval}}.
  \bibinfo{publisher}{ACM}, \bibinfo{address}{New York, NY, USA},
  \bibinfo{pages}{455–462}.
\newblock


\bibitem[\protect\citeauthoryear{Brown and et.al.}{Brown and et.al.}{2020}]%
        {Brown2020LanguageMA}
\bibfield{author}{\bibinfo{person}{T. Brown} {and} \bibinfo{person}{B.~Mann
  et.al.}} \bibinfo{year}{2020}\natexlab{}.
\newblock \showarticletitle{Language Models are Few-Shot Learners}.
\newblock \bibinfo{journal}{\emph{ArXiv}}  \bibinfo{volume}{abs/2005.14165}
  (\bibinfo{year}{2020}).
\newblock


\bibitem[\protect\citeauthoryear{Campos, Nguyen, Rosenberg, Song, Gao, Tiwary,
  Majumder, Deng, and Mitra}{Campos et~al\mbox{.}}{2016}]%
        {Campos2016MSMA}
\bibfield{author}{\bibinfo{person}{Daniel~Fernando Campos}, \bibinfo{person}{T.
  Nguyen}, \bibinfo{person}{M. Rosenberg}, \bibinfo{person}{Xia Song},
  \bibinfo{person}{Jianfeng Gao}, \bibinfo{person}{Saurabh Tiwary},
  \bibinfo{person}{Rangan Majumder}, \bibinfo{person}{L. Deng}, {and}
  \bibinfo{person}{Bhaskar Mitra}.} \bibinfo{year}{2016}\natexlab{}.
\newblock \showarticletitle{MS MARCO: A Human Generated MAchine Reading
  COmprehension Dataset}.
\newblock \bibinfo{journal}{\emph{ArXiv}}  \bibinfo{volume}{abs/1611.09268}
  (\bibinfo{year}{2016}).
\newblock


\bibitem[\protect\citeauthoryear{Chang, Yu, Chang, Yang, and Kumar}{Chang
  et~al\mbox{.}}{2019}]%
        {Chang2020PretrainingTF}
\bibfield{author}{\bibinfo{person}{Wei-Cheng Chang}, \bibinfo{person}{F. Yu},
  \bibinfo{person}{Yin-Wen Chang}, \bibinfo{person}{Yiming Yang}, {and}
  \bibinfo{person}{S. Kumar}.} \bibinfo{year}{2019}\natexlab{}.
\newblock \showarticletitle{Pre-training Tasks for Embedding-based Large-scale
  Retrieval}.
\newblock \bibinfo{journal}{\emph{International Conference on Learning
  Representations}} (\bibinfo{year}{2019}).
\newblock


\bibitem[\protect\citeauthoryear{Clarke, Craswell, and Soboroff}{Clarke
  et~al\mbox{.}}{2004}]%
        {Clarke2004OverviewOT}
\bibfield{author}{\bibinfo{person}{C. Clarke}, \bibinfo{person}{Nick Craswell},
  {and} \bibinfo{person}{I. Soboroff}.} \bibinfo{year}{2004}\natexlab{}.
\newblock \showarticletitle{Overview of the TREC 2004 Terabyte Track}. In
  \bibinfo{booktitle}{\emph{TREC}}.
\newblock


\bibitem[\protect\citeauthoryear{Clarke, Craswell, and Soboroff}{Clarke
  et~al\mbox{.}}{2009}]%
        {Clarke2009OverviewOT}
\bibfield{author}{\bibinfo{person}{C. Clarke}, \bibinfo{person}{Nick Craswell},
  {and} \bibinfo{person}{I. Soboroff}.} \bibinfo{year}{2009}\natexlab{}.
\newblock \showarticletitle{Overview of the TREC 2009 Web Track}. In
  \bibinfo{booktitle}{\emph{TREC}}.
\newblock


\bibitem[\protect\citeauthoryear{Craswell, Mitra, Yilmaz, Campos, and
  Voorhees}{Craswell et~al\mbox{.}}{2020}]%
        {Craswell2020OverviewOT}
\bibfield{author}{\bibinfo{person}{Nick Craswell}, \bibinfo{person}{Bhaskar
  Mitra}, \bibinfo{person}{E. Yilmaz}, \bibinfo{person}{Daniel~Fernando
  Campos}, {and} \bibinfo{person}{E. Voorhees}.}
  \bibinfo{year}{2020}\natexlab{}.
\newblock \showarticletitle{Overview of the TREC 2019 deep learning track}.
\newblock \bibinfo{journal}{\emph{ArXiv}}  \bibinfo{volume}{abs/2003.07820}
  (\bibinfo{year}{2020}).
\newblock


\bibitem[\protect\citeauthoryear{Dai and Callan}{Dai and Callan}{2019}]%
        {Dai2019DeeperTU}
\bibfield{author}{\bibinfo{person}{Zhuyun Dai} {and} \bibinfo{person}{J.
  Callan}.} \bibinfo{year}{2019}\natexlab{}.
\newblock \showarticletitle{Deeper Text Understanding for IR with Contextual
  Neural Language Modeling}.
\newblock \bibinfo{journal}{\emph{Proceedings of the 42nd International ACM
  SIGIR Conference on Research and Development in Information Retrieval}}
  (\bibinfo{year}{2019}).
\newblock


\bibitem[\protect\citeauthoryear{Dai, Xiong, Callan, and Liu}{Dai
  et~al\mbox{.}}{2018}]%
        {Dai2018ConvolutionalNN}
\bibfield{author}{\bibinfo{person}{Zhuyun Dai}, \bibinfo{person}{Chenyan
  Xiong}, \bibinfo{person}{J. Callan}, {and} \bibinfo{person}{Zhiyuan Liu}.}
  \bibinfo{year}{2018}\natexlab{}.
\newblock \showarticletitle{Convolutional Neural Networks for Soft-Matching
  N-Grams in Ad-hoc Search}.
\newblock \bibinfo{journal}{\emph{Proceedings of the Eleventh ACM International
  Conference on Web Search and Data Mining}} (\bibinfo{year}{2018}).
\newblock


\bibitem[\protect\citeauthoryear{Devlin, Chang, Lee, and Toutanova}{Devlin
  et~al\mbox{.}}{2019}]%
        {devlin2018bert}
\bibfield{author}{\bibinfo{person}{Jacob Devlin}, \bibinfo{person}{Ming-Wei
  Chang}, \bibinfo{person}{Kenton Lee}, {and} \bibinfo{person}{Kristina
  Toutanova}.} \bibinfo{year}{2019}\natexlab{}.
\newblock \showarticletitle{Bert: Pre-training of Deep Bidirectional
  Transformers for Language Understanding}. In
  \bibinfo{booktitle}{\emph{Proceedings of the 2019 Conference of the North
  {A}merican Chapter of the Association for Computational Linguistics: Human
  Language Technologies}}. \bibinfo{publisher}{The 58th Annual Meeting of the
  Association for Computational Linguistics}, \bibinfo{address}{Stroudsburg,
  PA, USA}, \bibinfo{pages}{4171--4186}.
\newblock


\bibitem[\protect\citeauthoryear{Fan, Guo, Lan, Xu, Zhai, and Cheng}{Fan
  et~al\mbox{.}}{2018}]%
        {Fan2018ModelingDR}
\bibfield{author}{\bibinfo{person}{Y. Fan}, \bibinfo{person}{J. Guo},
  \bibinfo{person}{Yanyan Lan}, \bibinfo{person}{J. Xu},
  \bibinfo{person}{ChengXiang Zhai}, {and} \bibinfo{person}{X. Cheng}.}
  \bibinfo{year}{2018}\natexlab{}.
\newblock \showarticletitle{Modeling Diverse Relevance Patterns in Ad-hoc
  Retrieval}.
\newblock \bibinfo{journal}{\emph{The 41st International ACM SIGIR Conference
  on Research and Development in Information Retrieval}}
  (\bibinfo{year}{2018}).
\newblock


\bibitem[\protect\citeauthoryear{Fan, Guo, Ma, Zhang, Lan, and Cheng}{Fan
  et~al\mbox{.}}{2021}]%
        {Fan2021}
\bibfield{author}{\bibinfo{person}{Yixing Fan}, \bibinfo{person}{Jiafeng Guo},
  \bibinfo{person}{Xinyu Ma}, \bibinfo{person}{Ruqing Zhang},
  \bibinfo{person}{Yanyan Lan}, {and} \bibinfo{person}{Xueqi Cheng}.}
  \bibinfo{year}{2021}\natexlab{}.
\newblock \showarticletitle{A Linguistic Study on Relevance Modeling in
  Information Retrieval}. In \bibinfo{booktitle}{\emph{Proceedings of the Web
  Conference 2021}}. \bibinfo{publisher}{{ACM}}.
\newblock
\urldef\tempurl%
\url{https://doi.org/10.1145/3442381.3450009}
\showDOI{\tempurl}


\bibitem[\protect\citeauthoryear{Gla{\ss}, Gliozzo, Chakravarti, Ferritto, Pan,
  Bhargav, Garg, and Sil}{Gla{\ss} et~al\mbox{.}}{2020}]%
        {Gla2020SpanSP}
\bibfield{author}{\bibinfo{person}{M. Gla{\ss}}, \bibinfo{person}{A. Gliozzo},
  \bibinfo{person}{Rishav Chakravarti}, \bibinfo{person}{Anthony Ferritto},
  \bibinfo{person}{Lin Pan}, \bibinfo{person}{G~P~Shrivatsa Bhargav},
  \bibinfo{person}{Dinesh Garg}, {and} \bibinfo{person}{A. Sil}.}
  \bibinfo{year}{2020}\natexlab{}.
\newblock \showarticletitle{Span Selection Pre-training for Question
  Answering}.
\newblock \bibinfo{journal}{\emph{The 58th Annual Meeting of the Association
  for Computational Linguistics}} (\bibinfo{year}{2020}).
\newblock


\bibitem[\protect\citeauthoryear{Guo, Fan, Ai, and Croft}{Guo
  et~al\mbox{.}}{2016}]%
        {Guo2016ADR}
\bibfield{author}{\bibinfo{person}{J. Guo}, \bibinfo{person}{Y. Fan},
  \bibinfo{person}{Qingyao Ai}, {and} \bibinfo{person}{W. Croft}.}
  \bibinfo{year}{2016}\natexlab{}.
\newblock \showarticletitle{A Deep Relevance Matching Model for Ad-hoc
  Retrieval}.
\newblock \bibinfo{journal}{\emph{Proceedings of the 25th ACM International on
  Conference on Information and Knowledge Management}} (\bibinfo{year}{2016}).
\newblock


\bibitem[\protect\citeauthoryear{Ke, Ji, Liu, Zhu, and Huang}{Ke
  et~al\mbox{.}}{2020}]%
        {Ke2020SentiLARESL}
\bibfield{author}{\bibinfo{person}{Pei Ke}, \bibinfo{person}{Haozhe Ji},
  \bibinfo{person}{Siyang Liu}, \bibinfo{person}{X. Zhu}, {and}
  \bibinfo{person}{Minlie Huang}.} \bibinfo{year}{2020}\natexlab{}.
\newblock \showarticletitle{SentiLARE: Sentiment-Aware Language Representation
  Learning with Linguistic Knowledge}. In \bibinfo{booktitle}{\emph{Proceedings
  of the 2020 Conference on Empirical Methods in Natural Language Processing}}.
\newblock


\bibitem[\protect\citeauthoryear{Kingma and Ba}{Kingma and Ba}{2015}]%
        {Kingma2015AdamAM}
\bibfield{author}{\bibinfo{person}{Diederik~P. Kingma} {and}
  \bibinfo{person}{Jimmy Ba}.} \bibinfo{year}{2015}\natexlab{}.
\newblock \showarticletitle{Adam: {A} Method for Stochastic Optimization}. In
  \bibinfo{booktitle}{\emph{3rd International Conference on Learning
  Representations, {ICLR} 2015, San Diego, CA, USA, May 7-9, 2015, Conference
  Track Proceedings}}, \bibfield{editor}{\bibinfo{person}{Yoshua Bengio} {and}
  \bibinfo{person}{Yann LeCun}} (Eds.).
\newblock


\bibitem[\protect\citeauthoryear{Lample, Ballesteros, Subramanian, Kawakami,
  and Dyer}{Lample et~al\mbox{.}}{2016}]%
        {lample-etal-2016-neural}
\bibfield{author}{\bibinfo{person}{Guillaume Lample}, \bibinfo{person}{Miguel
  Ballesteros}, \bibinfo{person}{Sandeep Subramanian}, \bibinfo{person}{Kazuya
  Kawakami}, {and} \bibinfo{person}{Chris Dyer}.}
  \bibinfo{year}{2016}\natexlab{}.
\newblock \showarticletitle{Neural Architectures for Named Entity Recognition}.
  In \bibinfo{booktitle}{\emph{Proceedings of the 2016 Conference of the North
  {A}merican Chapter of the Association for Computational Linguistics: Human
  Language Technologies}}. \bibinfo{publisher}{Association for Computational
  Linguistics}, \bibinfo{address}{San Diego, California},
  \bibinfo{pages}{260--270}.
\newblock


\bibitem[\protect\citeauthoryear{Lee, Chang, and Toutanova}{Lee
  et~al\mbox{.}}{2019}]%
        {Lee2019LatentRF}
\bibfield{author}{\bibinfo{person}{Kenton Lee}, \bibinfo{person}{Ming-Wei
  Chang}, {and} \bibinfo{person}{Kristina Toutanova}.}
  \bibinfo{year}{2019}\natexlab{}.
\newblock \showarticletitle{Latent Retrieval for Weakly Supervised Open Domain
  Question Answering}. In \bibinfo{booktitle}{\emph{Proceedings of the 57th
  Annual Meeting of the Association for Computational Linguistics}}.
  \bibinfo{publisher}{Association for Computational Linguistics},
  \bibinfo{address}{Florence, Italy}, \bibinfo{pages}{6086--6096}.
\newblock


\bibitem[\protect\citeauthoryear{Lin, Nogueira, and Yates}{Lin
  et~al\mbox{.}}{2020}]%
        {Lin2020PretrainedTF}
\bibfield{author}{\bibinfo{person}{Jimmy Lin}, \bibinfo{person}{Rodrigo
  Nogueira}, {and} \bibinfo{person}{A. Yates}.}
  \bibinfo{year}{2020}\natexlab{}.
\newblock \showarticletitle{Pretrained Transformers for Text Ranking: BERT and
  Beyond}.
\newblock \bibinfo{journal}{\emph{ArXiv}}  \bibinfo{volume}{abs/2010.06467}
  (\bibinfo{year}{2020}).
\newblock


\bibitem[\protect\citeauthoryear{Ma, Guo, Zhang, Fan, Ji, and Cheng}{Ma
  et~al\mbox{.}}{2020}]%
        {Ma2020PROPPW}
\bibfield{author}{\bibinfo{person}{Xinyu Ma}, \bibinfo{person}{Jiafeng Guo},
  \bibinfo{person}{Ruqing Zhang}, \bibinfo{person}{Yixing Fan},
  \bibinfo{person}{Xiang Ji}, {and} \bibinfo{person}{Xueqi Cheng}.}
  \bibinfo{year}{2020}\natexlab{}.
\newblock \showarticletitle{PROP: Pre-training with Representative Words
  Prediction for Ad-hoc Retrieval}.
\newblock \bibinfo{journal}{\emph{ArXiv}}  \bibinfo{volume}{abs/2010.10137}
  (\bibinfo{year}{2020}).
\newblock


\bibitem[\protect\citeauthoryear{MacAvaney, Yates, Cohan, and
  Goharian}{MacAvaney et~al\mbox{.}}{2019}]%
        {MacAvaney2019CEDRCE}
\bibfield{author}{\bibinfo{person}{Sean MacAvaney}, \bibinfo{person}{Andrew
  Yates}, \bibinfo{person}{Arman Cohan}, {and} \bibinfo{person}{Nazli
  Goharian}.} \bibinfo{year}{2019}\natexlab{}.
\newblock \showarticletitle{CEDR: Contextualized Embeddings for Document
  Ranking}.
\newblock \bibinfo{journal}{\emph{Proceedings of the 42nd International ACM
  SIGIR Conference on Research and Development in Information Retrieval}}
  (\bibinfo{year}{2019}).
\newblock


\bibitem[\protect\citeauthoryear{Nogueira and Cho}{Nogueira and Cho}{2019}]%
        {Nogueira2019PassageRW}
\bibfield{author}{\bibinfo{person}{Rodrigo Nogueira} {and}
  \bibinfo{person}{Kyunghyun Cho}.} \bibinfo{year}{2019}\natexlab{}.
\newblock \showarticletitle{Passage Re-ranking with BERT}.
\newblock \bibinfo{journal}{\emph{ArXiv}}  \bibinfo{volume}{abs/1901.04085}
  (\bibinfo{year}{2019}).
\newblock


\bibitem[\protect\citeauthoryear{Nogueira, Jiang, and Lin}{Nogueira
  et~al\mbox{.}}{2020}]%
        {Nogueira2020DocumentRW}
\bibfield{author}{\bibinfo{person}{Rodrigo Nogueira}, \bibinfo{person}{Zhiying
  Jiang}, {and} \bibinfo{person}{Jimmy Lin}.} \bibinfo{year}{2020}\natexlab{}.
\newblock \showarticletitle{Document Ranking with a Pretrained
  Sequence-to-Sequence Model}. In \bibinfo{booktitle}{\emph{Proceedings of the
  2020 Conference on Empirical Methods in Natural Language Processing}}.
\newblock


\bibitem[\protect\citeauthoryear{Nogueira, Yang, Lin, and Cho}{Nogueira
  et~al\mbox{.}}{2019}]%
        {Nogueira2019DocumentEB}
\bibfield{author}{\bibinfo{person}{Rodrigo Nogueira}, \bibinfo{person}{Wei
  Yang}, \bibinfo{person}{Jimmy Lin}, {and} \bibinfo{person}{Kyunghyun Cho}.}
  \bibinfo{year}{2019}\natexlab{}.
\newblock \showarticletitle{Document Expansion by Query Prediction}.
\newblock \bibinfo{journal}{\emph{ArXiv}}  \bibinfo{volume}{abs/1904.08375}
  (\bibinfo{year}{2019}).
\newblock


\bibitem[\protect\citeauthoryear{Pang, Lan, Guo, Xu, Xu, and Cheng}{Pang
  et~al\mbox{.}}{2017}]%
        {Pang2017DeepRankAN}
\bibfield{author}{\bibinfo{person}{Liang Pang}, \bibinfo{person}{Yanyan Lan},
  \bibinfo{person}{J. Guo}, \bibinfo{person}{Jun Xu}, \bibinfo{person}{J. Xu},
  {and} \bibinfo{person}{X. Cheng}.} \bibinfo{year}{2017}\natexlab{}.
\newblock \showarticletitle{DeepRank: A New Deep Architecture for Relevance
  Ranking in Information Retrieval}.
\newblock \bibinfo{journal}{\emph{Proceedings of the 2017 ACM on Conference on
  Information and Knowledge Management}} (\bibinfo{year}{2017}).
\newblock


\bibitem[\protect\citeauthoryear{Peters, Neumann, Iyyer, Gardner, Clark, Lee,
  and Zettlemoyer}{Peters et~al\mbox{.}}{2018}]%
        {peters-etal-2018-deep}
\bibfield{author}{\bibinfo{person}{Matthew Peters}, \bibinfo{person}{Mark
  Neumann}, \bibinfo{person}{Mohit Iyyer}, \bibinfo{person}{Matt Gardner},
  \bibinfo{person}{Christopher Clark}, \bibinfo{person}{Kenton Lee}, {and}
  \bibinfo{person}{Luke Zettlemoyer}.} \bibinfo{year}{2018}\natexlab{}.
\newblock \showarticletitle{Deep Contextualized Word Representations}. In
  \bibinfo{booktitle}{\emph{Proceedings of the 2018 Conference of the North
  {A}merican Chapter of the Association for Computational Linguistics: Human
  Language Technologies, Volume 1 (Long Papers)}}.
  \bibinfo{publisher}{Association for Computational Linguistics},
  \bibinfo{address}{New Orleans, Louisiana}, \bibinfo{pages}{2227--2237}.
\newblock


\bibitem[\protect\citeauthoryear{Qin and Liu}{Qin and Liu}{2013}]%
        {Qin2013IntroducingL4}
\bibfield{author}{\bibinfo{person}{T. Qin} {and} \bibinfo{person}{T. Liu}.}
  \bibinfo{year}{2013}\natexlab{}.
\newblock \showarticletitle{Introducing LETOR 4.0 Datasets}.
\newblock \bibinfo{journal}{\emph{ArXiv}}  \bibinfo{volume}{abs/1306.2597}
  (\bibinfo{year}{2013}).
\newblock


\bibitem[\protect\citeauthoryear{Raffel, Shazeer, Roberts, Lee, Narang, Matena,
  Zhou, Li, and Liu}{Raffel et~al\mbox{.}}{2020}]%
        {Raffel2020ExploringTL}
\bibfield{author}{\bibinfo{person}{Colin Raffel}, \bibinfo{person}{Noam
  Shazeer}, \bibinfo{person}{Adam Roberts}, \bibinfo{person}{Katherine Lee},
  \bibinfo{person}{Sharan Narang}, \bibinfo{person}{M. Matena},
  \bibinfo{person}{Yanqi Zhou}, \bibinfo{person}{W. Li}, {and}
  \bibinfo{person}{Peter~J. Liu}.} \bibinfo{year}{2020}\natexlab{}.
\newblock \showarticletitle{Exploring the Limits of Transfer Learning with a
  Unified Text-to-Text Transformer}.
\newblock \bibinfo{journal}{\emph{J. Mach. Learn. Res.}}  \bibinfo{volume}{21}
  (\bibinfo{year}{2020}), \bibinfo{pages}{140:1--140:67}.
\newblock


\bibitem[\protect\citeauthoryear{Rajpurkar, Zhang, Lopyrev, and
  Liang}{Rajpurkar et~al\mbox{.}}{2016}]%
        {Rajpurkar2016SQuAD10}
\bibfield{author}{\bibinfo{person}{Pranav Rajpurkar}, \bibinfo{person}{Jian
  Zhang}, \bibinfo{person}{Konstantin Lopyrev}, {and} \bibinfo{person}{Percy
  Liang}.} \bibinfo{year}{2016}\natexlab{}.
\newblock \showarticletitle{SQuAD: 100, 000+ Questions for Machine
  Comprehension of Text}. In \bibinfo{booktitle}{\emph{Proceedings of the 2016
  Conference on Empirical Methods in Natural Language Processing}}.
\newblock


\bibitem[\protect\citeauthoryear{Robertson and Zaragoza}{Robertson and
  Zaragoza}{2009}]%
        {Robertson2009ThePR}
\bibfield{author}{\bibinfo{person}{S. Robertson} {and} \bibinfo{person}{H.
  Zaragoza}.} \bibinfo{year}{2009}\natexlab{}.
\newblock \showarticletitle{The Probabilistic Relevance Framework: BM25 and
  Beyond}.
\newblock \bibinfo{journal}{\emph{Found. Trends Inf. Retr.}}
  \bibinfo{volume}{3} (\bibinfo{year}{2009}), \bibinfo{pages}{333--389}.
\newblock


\bibitem[\protect\citeauthoryear{Vaswani, Shazeer, Parmar, Uszkoreit, Jones,
  Gomez, Kaiser, and Polosukhin}{Vaswani et~al\mbox{.}}{2017}]%
        {Vaswani2017AttentionIA}
\bibfield{author}{\bibinfo{person}{Ashish Vaswani}, \bibinfo{person}{Noam
  Shazeer}, \bibinfo{person}{Niki Parmar}, \bibinfo{person}{Jakob Uszkoreit},
  \bibinfo{person}{Llion Jones}, \bibinfo{person}{Aidan~N. Gomez},
  \bibinfo{person}{undefinedukasz Kaiser}, {and} \bibinfo{person}{Illia
  Polosukhin}.} \bibinfo{year}{2017}\natexlab{}.
\newblock \showarticletitle{Attention is All You Need}. In
  \bibinfo{booktitle}{\emph{Proceedings of the 31st International Conference on
  Neural Information Processing Systems}} (Long Beach, California, USA).
  \bibinfo{publisher}{Curran Associates Inc.}, \bibinfo{address}{Red Hook, NY,
  USA}, \bibinfo{pages}{6000–6010}.
\newblock


\bibitem[\protect\citeauthoryear{Voorhees}{Voorhees}{2004}]%
        {Voorhees2004OverviewOT}
\bibfield{author}{\bibinfo{person}{E. Voorhees}.}
  \bibinfo{year}{2004}\natexlab{}.
\newblock \showarticletitle{Overview of the TREC 2004 Robust Track.}
\newblock


\bibitem[\protect\citeauthoryear{Wang, Singh, Michael, Hill, Levy, and
  Bowman}{Wang et~al\mbox{.}}{2018}]%
        {Wang2018GLUEAM}
\bibfield{author}{\bibinfo{person}{Alex Wang}, \bibinfo{person}{Amanpreet
  Singh}, \bibinfo{person}{Julian Michael}, \bibinfo{person}{Felix Hill},
  \bibinfo{person}{Omer Levy}, {and} \bibinfo{person}{Samuel~R. Bowman}.}
  \bibinfo{year}{2018}\natexlab{}.
\newblock \showarticletitle{GLUE: A Multi-Task Benchmark and Analysis Platform
  for Natural Language Understanding}. In
  \bibinfo{booktitle}{\emph{BlackboxNLP@EMNLP}}.
\newblock


\bibitem[\protect\citeauthoryear{Yang, Zhang, and Lin}{Yang
  et~al\mbox{.}}{2019b}]%
        {Yang2019SimpleAO}
\bibfield{author}{\bibinfo{person}{Wei Yang}, \bibinfo{person}{Haotian Zhang},
  {and} \bibinfo{person}{Jimmy Lin}.} \bibinfo{year}{2019}\natexlab{b}.
\newblock \showarticletitle{Simple Applications of BERT for Ad Hoc Document
  Retrieval}.
\newblock \bibinfo{journal}{\emph{ArXiv}}  \bibinfo{volume}{abs/1903.10972}
  (\bibinfo{year}{2019}).
\newblock


\bibitem[\protect\citeauthoryear{Yang, Dai, Yang, Carbonell, Salakhutdinov, and
  Le}{Yang et~al\mbox{.}}{2019a}]%
        {Yang2019XLNetGA}
\bibfield{author}{\bibinfo{person}{Z. Yang}, \bibinfo{person}{Zihang Dai},
  \bibinfo{person}{Yiming Yang}, \bibinfo{person}{J. Carbonell},
  \bibinfo{person}{R. Salakhutdinov}, {and} \bibinfo{person}{Quoc~V. Le}.}
  \bibinfo{year}{2019}\natexlab{a}.
\newblock \showarticletitle{XLNet: Generalized Autoregressive Pretraining for
  Language Understanding}. In \bibinfo{booktitle}{\emph{NeurIPS}}.
\newblock


\bibitem[\protect\citeauthoryear{Zhai}{Zhai}{2008}]%
        {Zhai2008StatisticalLM}
\bibfield{author}{\bibinfo{person}{ChengXiang Zhai}.}
  \bibinfo{year}{2008}\natexlab{}.
\newblock \showarticletitle{Statistical Language Models for Information
  Retrieval: A Critical Review}.
\newblock \bibinfo{journal}{\emph{Found. Trends Inf. Retr.}}
  \bibinfo{volume}{2} (\bibinfo{year}{2008}), \bibinfo{pages}{137--213}.
\newblock


\bibitem[\protect\citeauthoryear{Zhang, Zhao, Saleh, and Liu}{Zhang
  et~al\mbox{.}}{2020}]%
        {zhang2019pegasus}
\bibfield{author}{\bibinfo{person}{Jingqing Zhang}, \bibinfo{person}{Yao Zhao},
  \bibinfo{person}{Mohammad Saleh}, {and} \bibinfo{person}{Peter Liu}.}
  \bibinfo{year}{2020}\natexlab{}.
\newblock \showarticletitle{{PEGASUS}: Pre-training with Extracted
  Gap-sentences for Abstractive Summarization}. In
  \bibinfo{booktitle}{\emph{Proceedings of the 37th International Conference on
  Machine Learning}}, \bibfield{editor}{\bibinfo{person}{Hal~Daumé III} {and}
  \bibinfo{person}{Aarti Singh}} (Eds.). \bibinfo{pages}{11328--11339}.
\newblock


\end{thebibliography}
